# Combinatorial Development of Amorphous/nanocrystalline Biphase Soft Magnetic Alloys with Silicon-steel like Saturated Magnetic Induction

Xuesong Li[1,2 #], Jing Zhou[1,2 #], Xiao Liu[1], Xibei Hou[1], Tengyu Guo[1], Bo Wu[3], Baoan Sun[1,2,4 *], Weihua Wang[1,2,4], Haiyang Bai[1,2,4 *]

[1]*Songshan Lake Materials Laboratory, Dongguan 523808, China*
[2]*Institute of Physics, Chinese Academy of Sciences, Beijing 100190, China*
[3]*Bay Area Center of Electron Microscopy, Dongguan 523808, China*
[4]*Center of Materials Science and Optoelectronics Engineering, University of Chinese Academy of Sciences, Beijing 100049, China*

[#]These authors contributed equally to this work.

[*]Corresponding authors: sunba@iphy.ac.cn (B. A. Sun); hybai@iphy.ac.cn (H.Y. Bai)

**Abstract**

Maximization saturation magnetic induction ($B_s$) of soft magnetic alloys is essential for the high power-density electromagnetic devices. However, identifying the alloy compositions with high $B_s$ often replies on the lab-intensive melt casting method and a high-throughput characterization on magnetic properties remains challenging. Here, we develop a new combinatorial method for fast screening alloys with optimal soft magnetic properties based on the high-throughput MOKE screening method. Based on the combinatorial method, we found that the alloys with a combination of high $B_s$ and low coercivity ($H_c$) tend to have a feature of amorphous-nanocrystalline biphase microstructure. We also identified an amorphous/nanocrystalline alloy film with the composition the $Fe_{68.09}Co_{17.02}B_{10.9}Si_4$, exhibiting an ultra-high $B_s$ up to 2.02 T that surpasses all amorphous/nanocrystalline alloys reported so far and is comparable to that of silicon steels, together with a high resistivity of 882 μΩ·cm, about 17 times of silicon steels. Our high-throughput magnetic screening method provides a paradigm for understanding the relationship between microstructure and magnetic properties and the development of the next-generation soft magnetic materials.

# Introduction

Since Michael Faraday's discovery of electromagnetic induction [1], soft magnetic materials have been an indispensable component of modern electronics and play a key role in the transmission, conversion, and storage of electro-magnetic energy [2, 3]. Over the past century, a variety of soft magnetic materials such as silicon steels, permalloys, soft ferrite, to name but a few, with different compositions and microscopic structures have been subsequently developed [4-9]. Among these materials, amorphous alloys (mainly based on iron) and their descendent nanocrystalline alloys, are considered as the next-generation soft magnetic materials owing to their low $H_c$ and energy losses, particularly for high-frequency applications [10]. However, in order to retain the amorphous structure, there is often a limitation for the content of magnetic elements in these alloys. This leads to a relatively low $B_s$ as compared to other soft magnetic alloys. For example, the $B_s$ of commercial amorphous alloy (METGLAS) is ~1.56 T [11], which is much lower than that of silicon steels (1.9-2.0 T) [12]. Increasing the saturation magnetic induction has become an urgent task for the amorphous/nanocrystalline alloys (ANAs). However, the ANAs are typically multiple-element alloys that contains at least three elements [13]. Development of high-performance soft magnetic ANAs is still based on the conventional try-and-error approach, which appears to be low-efficiency and labor-intensive. Exploring the optimal material among millions of possible combinations of chemical composition and microstructure remains challenging.

In recent years, materials discovery and design have taken a giant leap with the applications of combinatorial approach in different areas [14, 15]. By combining the efficient synthesis of the combinatorial library with the high-throughput property-screening method, the composition-microstructure-property relationship can be delineated, and hence compositions with desired properties are rapidly identified. At present, a series of high-throughput screening technologies based on different core indicators, such as resistivity, elastic modulus, thermal conductivity, diffraction peak strength, etc., have been developed [16-21], and a series of structural and functional materials with excellent properties were also successfully developed. However, there is rare precedent for such efficient methods to screen soft magnetic material. Traditional methods of directly measuring soft magnetic properties, such as the $B_s$ and the $H_c$, are highly sensitive to the shape, magnetization state and direction, and the stress distribution in the material [22]. This makes it extremely difficult to directly perform high-throughput screening on core indicators of soft magnetic materials.

Advanced surface magneto-optical Kerr effect (MOKE) technology provides the possibility for this route [23]. The polarization plane of linearly polarized light rotates as the light is transmitted through a magnetized medium, and the magnitude of the MOKE signal depends on the soft magnetic properties. In this work, we successfully realized the high-throughput screening on magnetic properties in thin-film material library through the advanced surface MOKE technology. Based on the technology, the correlation between MOKE signals and soft magnetic properties for the composition library was established. A soft magnetic film composition of $Fe_{68.09}Co_{17.02}B_{10.9}Si_4$ was successfully developed, exhibiting an unprecedently high $B_s$ in ANAs that is comparable to silicon steels together with an ultra-high resistivity of 882 μΩ·cm. These excellent properties of the alloy are found to result from the amorphous-nanocrystalline biphase microstructure, which maximizes the average atomic magnetic moment and complicates the electronic structure.

## Results and discussion

### Fabrication and structural characterizations of combinatorial alloy library

It is known that fully crystal structure makes Fe-Si soft magnetic alloys exhibit high $B_s$ value of 1.9~2.0 T [12]. However, the concomitant high $H_c$ as well as high eddy current loss caused by the low resistivity of crystal structure result in huge energy dissipation in practical application. Amorphization is an effective method to increase the resistivity with less $B_s$ sacrifice, pathing by alloying an appropriate amount of B element in Fe-Si alloy systems [24]. Based on our previous study, the soft magnetic alloys should exhibit the largest local atomic magnetic moment when the ratio of Fe to Co is around 4 to 1, which is conducive to improving the $B_s$ [25]. Therefore, we simultaneously fabricated a large component library by three-target magnetron co-sputtering with chemical compositions of $Fe_8Co_2$ alloy, pure Si and pure B, respectively (see *Methods* for more experimental details). Figure 1a shows the schematic diagram of co-sputtering three-target magnetron co-sputtering method. The range of alloys required for the composition in the library can be realized by changing the orientation angle of the sputtering target relative to the substrate and adjusting the sputtering power of the target. As shown in Figure 1b, the composition libraries with hundreds of ingredients were deposited on a silicon substrate with a diameter of 100 mm from targets. Given the significant impact of the content of ferromagnetic elements on the $B_s$ of magnetic materials, we deliberately increased the sputtering power of the $Fe_8Co_2$ target in the fabrication of the combinatorial library. Here, we selected a sputtering power ratio of 185:15:8 for $Fe_8Co_2$, B, and Si targets. The substrate position remained invariant and the sputtering time was fixed at 1 hour throughout the entire process, resulting in a combinatorial library with a thickness of approximately 1 μm.

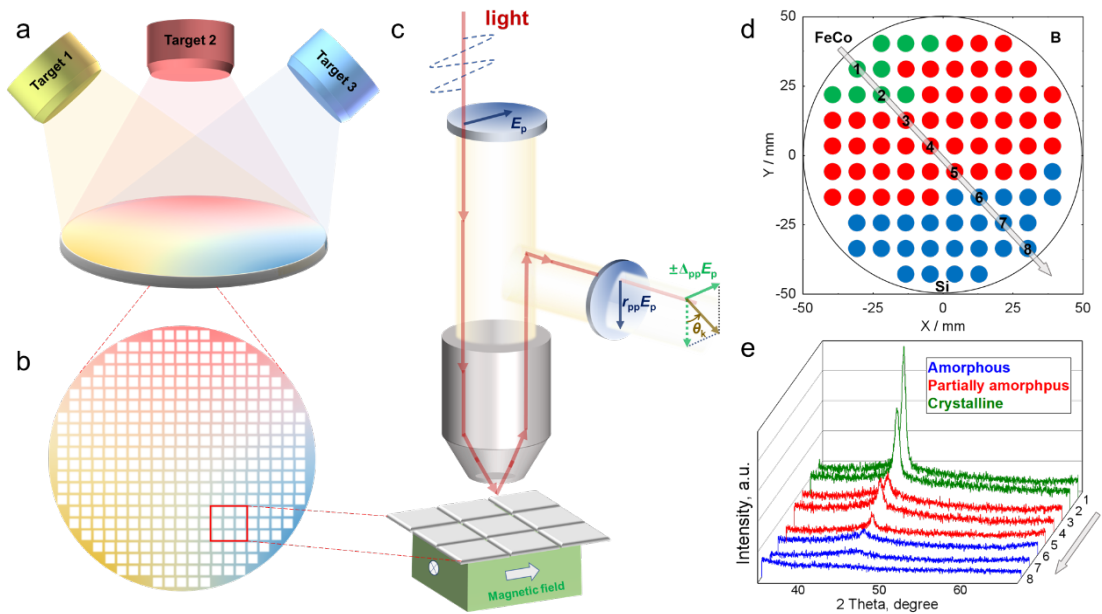

**Figure 1. The paradigm for the design of high $B_s$ SMMs based on high-throughput MOKE technology.** (a) Schematic diagram of three-target magnetron co-sputtering. (b) Schematic diagram and physical drawings of the combinatorial library deposited on a 100-mm-diameter silicon wafer. The composition reservoir area is divided into plenty of equal area small regions. (c) Schematic diagram of a measurement method using a MOKE microscope. (d) Structure map of the combinatorial library. (e) Corresponding XRD diffractograms showing the transition from crystalline to amorphous along one composition line, as indicated by the arrow in (d).

The X-ray diffraction (XRD) experiments were conducted to characterize the structure of the combinatorial library, where the spacing between X-ray spots in the horizontal and vertical directions was 10 mm (Figure S1). Figures 1d and 1e shows the structure map of the combinatorial library and corresponding XRD diffractograms. The entire combinatorial library is mainly composed of three structural types, namely amorphous (blue), partial amorphous (red) and crystalline (green), which have been distinguished by color in Figure 1d. The situation of each target was marked in the directions of the map, representing the approximate position of each target in the preparation process of the component library. The farther away from the target, the lower the content of the element in this region. As indicated by the arrow in Figure 1d, the corresponding XRD diffractograms show the transition from crystalline to glass structure along the direction of $Fe_8Co_2$ target (Figure 1e), which were labeled with serial numbers 1 to 8. Only crystalline structure can be formed even at such a high quenching rate of magnetron sputtering method when the proportion of ferromagnetic elements is too high. With decreasing the proportion of ferromagnetic elements, the glass forming ability of the region is improved under the induction of such metalloid elements as Si and B, and the region gradually transitions to a partially amorphous structure and eventually form a completely amorphous structure [24].

**High-throughput characterization of magnetic properties**

Magneto-optical effect is known that the polarization plane of linearly polarized light rotates as the light is transmitted through a magnetized medium. The effect of the magnetized medium was to rotate the polarization plane of the incident light by an angle ($\theta$) that depended on the strength of the magnetic field. Under the conditions holding for visible light, the effect of magnetic field on light is given by [23]:

$$\theta = \frac{2\pi n e^3 \omega^2 L(H+M)}{m^2 c^2 \omega_0^4 \sqrt{1+4\pi n e^2/m\omega_0^2}} \quad (1)$$

where $n$ is the electron density, $L$ is the length of sample, $H$ is the external applied magnetic field, $M$ is the magnetization of magnetized medium, $\omega$ is the light frequency. $e$, $c$ and $\omega_0$ are constants, representing elementary charge, light speed, and the natural frequency, respectively. It follows from Equation (1) that $\theta$ is positively correlated to

*M*.

Since metallic magnetic materials absorb light strongly, it is general to measure the reflected (Kerr effect) than the transmitted signal to probe the magneto-optic effect. As for surface MOKE technique, when we consider only linear p-polarized lasers as the light source for discussion, the Kerr intensity (*I*) measured by the photodiode after the light has passed through the analyzing polarizer is given by $I = |E_p|^2|\delta + \theta' + \theta''|^2 \approx |E_p|^2(\delta^2 + 2\delta\theta')^2 = |E_p|^2\delta^2\left(1 + \frac{2\theta'}{\delta}\right)$, where $\theta'$ and $\theta''$ represent Kerr rotation and the Kerr ellipticity respectively, and $\delta$ is about 1°-2° from the extinction condition to provide a DC bias in measurement. Now that the Kerr rotation is the change in light caused by the magneto-optic effect, the $\theta$ in Equation (1) can be treated as $\theta'$. Then the measured Kerr intensity is about [26]:

$$I = I_0\left(1 + 2\frac{2\pi n e^3 \omega^2 L(H+M)}{m^2 c^2 \omega_0^4 \delta \sqrt{1+4\pi n e^2/m\omega_0^2}}\right) \quad (2)$$

When the magnetic field *H* is fixed at a constant value $H_0$ sufficient to magnetize the magnetic medium to saturation, the corresponding magnetization *M* is the saturation magnetization $M_s$. Substituting these parameters into Equation (2) and simplifying yields the linear relation as follows:

$$I = aM_s + b \quad (3)$$

where *a* represents $\frac{4\pi n e^3 \omega^2 L}{m^2 c^2 \omega_0^4 \delta \sqrt{1+4\pi n e^2/m\omega_0^2}}$ and *b* represent $I_0 + \frac{4\pi n e^3 \omega^2 LH}{m^2 c^2 \omega_0^4 \delta \sqrt{1+4\pi n e^2/m\omega_0^2}}$.

Considering only variations in the magnetized medium, both *a* and *b* are constants, with $a > 0$. As evident from Equation (3), *I* increase linearly with the $M_s$ of magnetic medium. Therefore, it is theoretically feasible to analyze the $M_s$ by measuring the variation pattern of the Kerr intensity, as long as the applied magnetic field is identical and the magnetic medium has been magnetized to saturation.

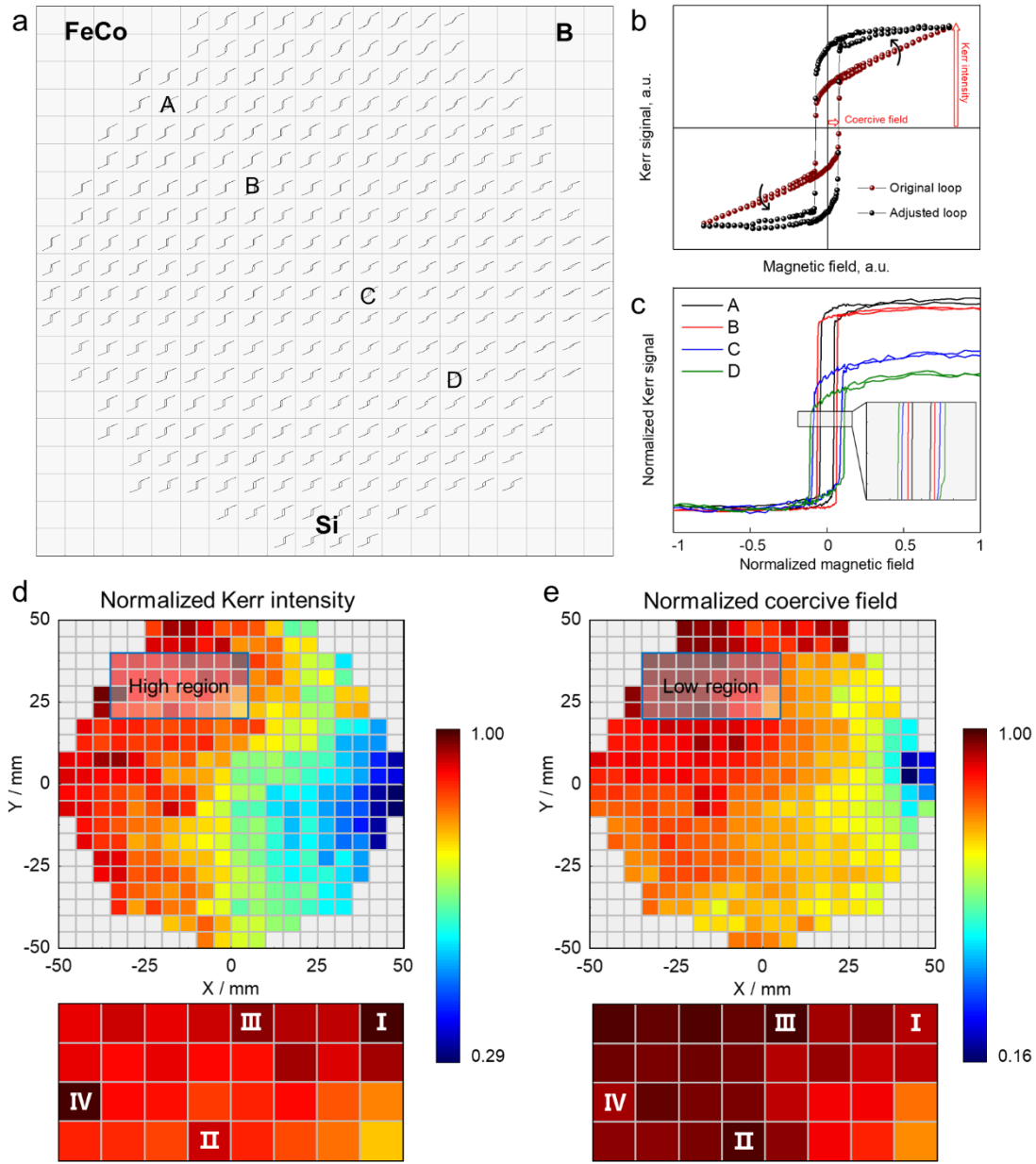

**Figure 2. Magnetic properties' maps of the FeCo-B-Si combinatorial library using a MOKE microscope.** (a) The Kerr hysteresis loops in combinatorial library, where the spacing between test points in the horizontal and vertical directions was 5 mm to ensure efficiency and accuracy. (b) Diagram of the drift and farad corrections method of Kerr hysteresis loop. (c) Corresponding Kerr hysteresis loops showing the variations in soft magnetic properties of the regions coded A. B. C and D, as indicated by the symbol in (a). Summary of the normalized MOKE intensities (d) and coercive fields (e) for all regions to the positions.

Based on the above analysis, the magnetic properties of the combinatorial library were further characterized using a high-throughput MOKE screening method. The spacing between test points in the horizontal and vertical directions was 5 mm to ensure efficiency and accuracy, dividing the component reservoir into near 300 separate regions. The Kerr hysteresis loop with magnetization direction parallel to the plane of combinatorial library is measured for each small region (Figure 1c). It should be noted that the amplitude of the applied magnetic field strength needs to be consistent. All of the Kerr hysteresis loops in combinatorial library are shown in Figure 2a (original data listed in Figure S2). In practice, the original Kerr loop measured using a MOKE microscope is usually inclined or non-closed, not in the classical form of hysteresis loop. Therefore, we uniformly adjusted the Kerr hysteresis loops using drift and farad method, so that the saturated part of the loop was uniformly parallel to the horizontal axis (Figure 2b). Then we extract the key parameters from the loop, that is, the intercept from the intersection with the horizontal axis to the origin corresponding to the coercive field, and the distance from the highest point of the loop to the horizontal axis corresponding to the Kerr intensity. Since the measurement process is continuous, we kept the amplitude of the magnetic field intensity applied during the magnetization process of all small regions to be consistent in order to ensure that the Kerr intensities are comparable. After normalization processing, we can get the relative soft magnetic properties of different regions in the component library, and finally screen out the best soft magnetic properties' region with both high $B_s$ and low $H_c$. Figure 2c shows the corrected Kerr hysteresis loops after normalization process for regions coded A, B, C and D of Figure 2a. One can see that the Kerr intensity decreases as the proportion of ferromagnetic elements decreases because the average atomic magnetic moment is diluted by the non-magnetic elements [27]. The same tendency is also found in coercive fields, which seems to be contrary to conventional cognition. The coercivity field in the amorphous region is higher than that in the crystalline region. It is speculated that this phenomenon may arise from the variation in the energy and distribution of atomic clusters during the deposition process caused by the combinatorial component library fabrication method [28], thus affecting the local magnetic anisotropy.

We summarized the normalized Kerr intensities and coercive fields of all corresponding small regions, forming the distribution maps shown in Figures 2d and

2e, respectively. Both the high Kerr intensity and the low coercive fields are located near the $Fe_8Co_2$ target. It is also found that the coercive field near B target is higher compared with that of Si target, indicating that Si element is helpful for softening soft magnetic materials. In order to identify soft magnetic materials with high $B_s$ and low $H_c$, alloy systems that are rich ferromagnetic elements within the region of low coercive field (see the thin solid line in Figure 2d) are preferentially selected. Four locations representing both high Kerr intensity and low coercive field in the region were identified, whose chemical composition and structure made them exhibit high $B_s$ and low loss were code-named I, II, III and IV, respectively. The element contents were measured using inductively coupled plasma emission spectroscopy (ICP-OES) to accurately detect the average chemical composition of these potentially excellent soft magnetic properties. In order to avoid the error of high silicon content in the film composition caused by the dissolution of the silicon substrate during the measurement process, we selected single crystal copper as the composition library of the sputtering substrate for composition measurement. Taking the average of the three measurement values, we identified the chemical compositions of locations I to IV as $Fe_{67.6}Co_{16.9}B_{11.5}Si_4$, $Fe_{68.09}Co_{17.02}B_{10.9}Si_4$, $Fe_{68.4}Co_{17.1}B_{10.9}Si_{3.6}$ and $Fe_{68.56}Co_{17.14}B_{11}Si_{3.3}$, respectively. It can be inferred from Figure 1d that the structures corresponding to these positions are located at the junction of pure crystal and partially amorphous region, indicating that the optimal magnetic properties in this alloy system generally exhibit an amorphous-nanocrystalline biphase structure. In addition, although the microstructure of the component library is not directly responded to the properties of magnetic materials, its decisive effect on the resistivity will affect the energy loss of the material from the side [29]. Therefore, considering only soft magnetic properties is unilateral, these compositions with a high proportion of amorphous structures are also beneficial to maximize the resistivity of magnetic material.

**Development of alloy films with unprecedented magnetic properties**

Independent alloy targets corresponding to each chemical composition were produced of locations I to IV in the combinatorial library, and uniform thin films were fabricated using a single target magnetron sputtering method. The rotation speed of the substrate during the sputtering process is 5 r/min. In order to eliminate the influence of thickness on the structure of the film, the sputtering power and time are adjusted to ensure that the thickness of the films are roughly the same as that of the component library. Figure 3a shows the XRD patterns of code-named I to IV thin films. It is clear that all of films show features of amorphous-nanocrystalline biphase structure. From the diffraction peak near 45°, it can be judged that the precipitated nanocrystal is *bcc*-Fe and the intensity change of diffraction peak is consistent with the result in Figure 1d. Figure 3b shows the hysteresis loops of as-quenched thin film measured by a magnetic property measurement system. All films were gradually magnetized to saturation as the applied magnetic field intensity increasing to 1500 kA/m. Among them, $Fe_{68.09}Co_{17.02}B_{10.9}Si_4$ film shows the highest $B_s$ up to 2.02 T. To the best of our knowledge, this $B_s$ value well exceeds the highest value of $B_s$ (~1.94 T) [25] among all ANAs reported previously. The $B_s$ value for the alloy is even comparable or surpass that of silicon steels and pure iron. Interestingly, the $B_s$ of the film fabricated by magnetron sputtering method is not completely proportional to the content of ferromagnetic elements. Excess ferromagnetic element content reduces the interatomic distance and increases the coordination number, resulting in the decrease of the average magnetic moment [30, 31].

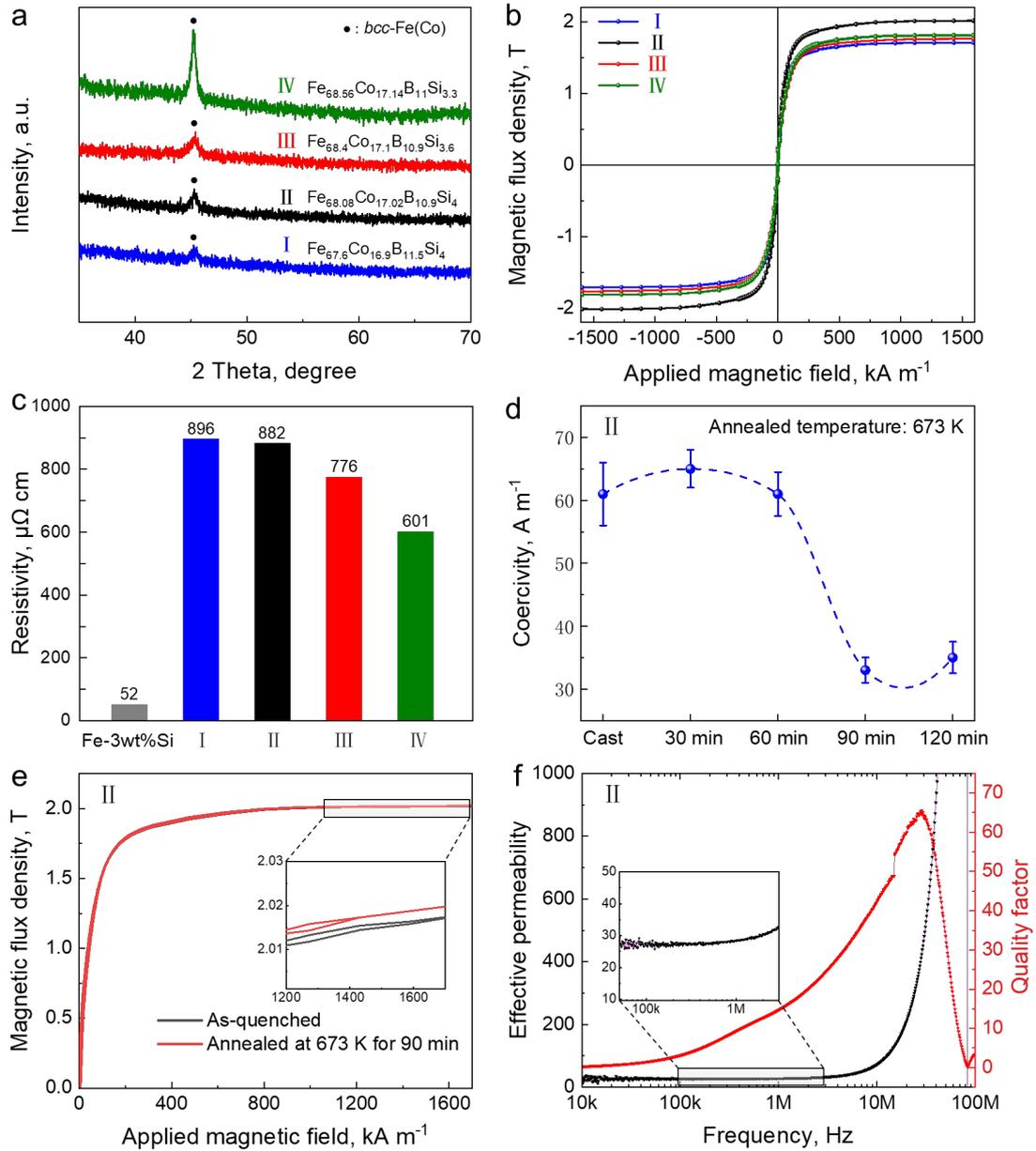

**Figure 4. Properties of the selected FeCo-B-Si alloy films.** (a) The XRD patterns of code-named I to IV thin films. The chemical compositions of locations I to IV were $Fe_{67.6}Co_{16.9}B_{11.5}Si_4$, $Fe_{68.08}Co_{17.02}B_{10.9}Si_4$, $Fe_{68.4}Co_{17.1}B_{10.9}Si_{3.6}$ and $Fe_{68.56}Co_{17.14}B_{11}Si_{3.3}$, respectively. (b) The hysteresis loops of code-named I to IV as-quenched thin films measured by VSM. (c) The resistivity of thin films I-IV and Fe-3wt%Si alloys measured by a standard four-probe method. (d) Changes in $H_c$ with isothermal time at 673 K of $Fe_{68.08}Co_{17.02}B_{10.9}Si_4$ soft magnetic film. (e) The $B_s$ of the $Fe_{68.08}Co_{17.02}B_{10.9}Si_4$ soft magnetic film after isothermal 90 min at 673 K compared with that of as-quenched state. (f) The frequency variation curve of effective permeability and quality factor of $Fe_{68.08}Co_{17.02}B_{10.9}Si_4$ film isothermal at 673 K for 90 min.

In addition to $B_s$, the resistivity is also an important parameter for the magnetic films in practical applications. We also measured the resistivity of thin films I-IV compared with Fe-3wt%Si alloy using a standard four-probe method, as shown in Figure 3c. It is clear that the resistivity of all films is much higher than that of the crystalline Fe-3wt%Si alloy, and increases with the increase of the volume fraction of amorphous phase. $Fe_{68.09}Co_{17.02}B_{10.9}Si_4$ soft magnetic film also exhibits high resistivity up to 882 μΩ·cm, about 17 times that of Fe-3wt%Si alloys. Compared with the crystalline structure, the atomic arrangement of the amorphous structure is irregular and its electronic structure is more complex. When the electrons are oriented, they are more susceptible to obstruction and thus exhibit higher resistivity. We also carried out a series of annealing processes on these alloy films. $Fe_{68.09}Co_{17.02}B_{10.9}Si_4$ soft magnetic film were isothermal at 673 K for various time under a vacuum annealing furnace. As shown in Figure 3d, when the annealing time is 90 min, the $H_c$ decreases from 61 A/m in the as-cast state to 33 A/m. We also measured the $B_s$ under the optimal annealing process (annealing time: 90 min), which was comparable to its as-quenched state within the error range (Figure 3e). Figure 3f shows the frequency variation curve of effective permeability and quality factor of $Fe_{68.09}Co_{17.02}B_{10.9}Si_4$ film isothermal at 673 K for 90 min. The effective permeability hardly decays from 10 kHz to 5 MHz and remains constant around 27. The frequency when the quality factor reaches the peak is 30 MHz. Such a high cut-off frequency means that this kind of soft magnetic material still maintain its magnetic characteristics at frequencies up to 30 MHz [32].

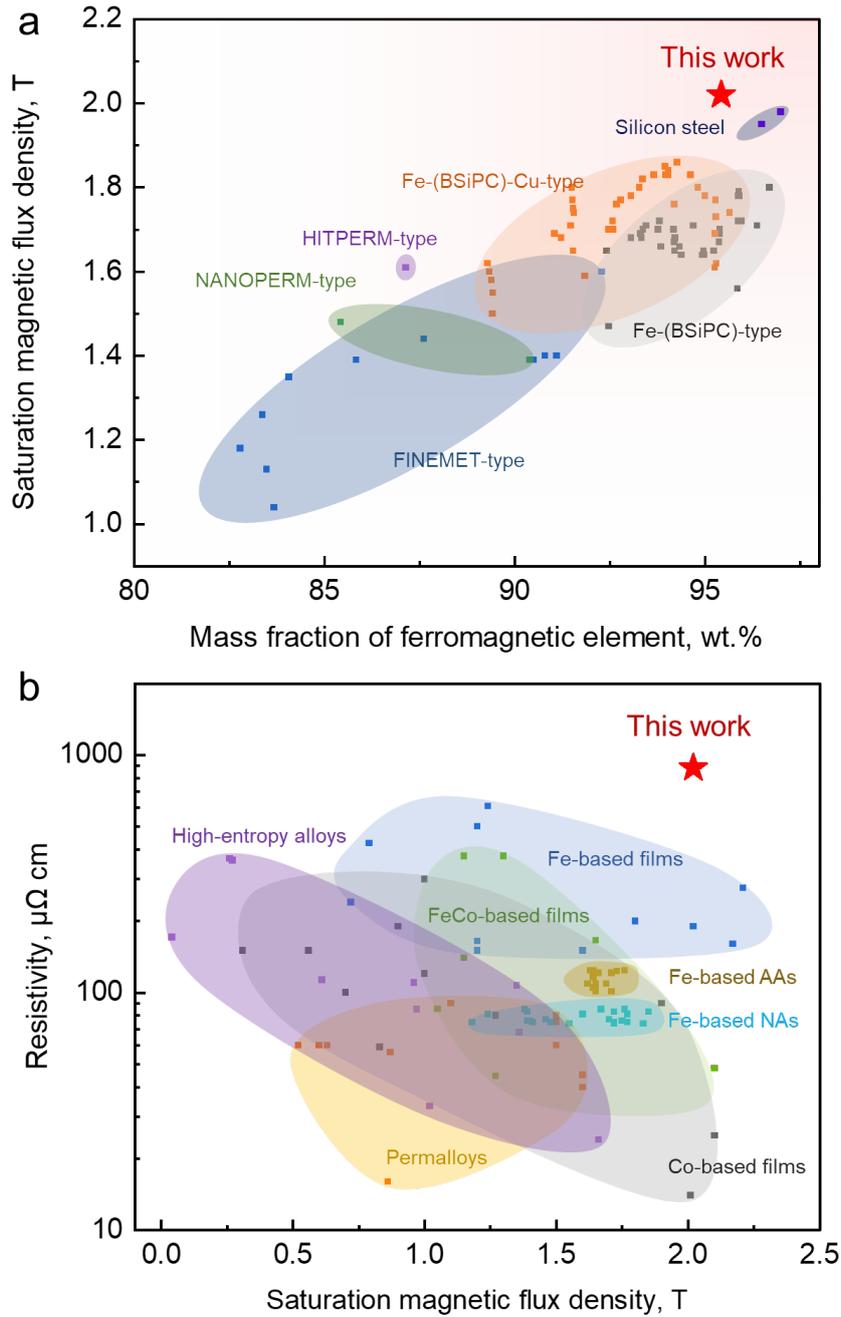

**Figure 4. Summary of the magnetic properties (e.g., $B_s$, $H_c$ and resistivity) of various magnetic materials.** (a) Variation of $B_s$ with the mass fraction of ferromagnetic element for previously reported with Fe-based amorphous-nanocrystalline biphase structures and silicon steels. (b) Relationship between the $B_s$ and resistivity of the typical soft magnetic alloys and magnetic alloy films.

To highlight the excellent comprehensive properties of $Fe_{68.09}Co_{17.02}B_{10.9}Si_4$ magnetic film, we have collected the existing data of magnetic films (Fe-based [33-39], Co-based [40-45] and FeCo-based [46-48]), Fe-based amorphous alloys (AAs) [49-52],

Fe-based nanocrystalline alloys (NAs) [9, 53-73], Silicon steel [74], permalloys [75, 76] and high-entropy alloys [77-83]. We compared the $B_s$ value of the alloys previously reported with Fe-based amorphous-nanocrystalline biphase structures with the mass fraction of ferromagnetic elements (Figure 4a). Intriguingly, the total content of ferromagnetic element in $Fe_{68.09}Co_{17.02}B_{10.9}Si_4$ magnetic film is not large enough, but reaches the maximum $B_s$ value of 2.02 T (comparable to the limit value of silicon-steels). Figure 4b shows the relationship between the $B_s$ and resistivity of the typical soft magnetic alloys and magnetic alloy films. As can be seen, the increment of $B_s$ usually accompanied by the deterioration of resistivity. Here, the $B_s$ and resistivity value of $Fe_{68.09}Co_{17.02}B_{10.9}Si_4$ magnetic film develop by high-throughput MOKE screening method reaches 2.02 T and 882 $\mu\Omega \cdot cm$, respectively, realizing a combination of the high $B_s$ and resistivity among all soft magnetic alloys and magnetic films. We have also summarized the changes in $B_s$ with $H_c$ of high-entropy alloys and magnetic alloy films, as shown in Figure S3. As for high-entropy alloys and all types of magnetic films, although a small number of components' $B_s$ exceeds 1.95 T, the $H_c$ of all components is above 30 A/m, and the maximum value even exceeds 3000 A/m. As for Fe-based amorphous alloys, Fe-based nanocrystalline alloys, and permalloys, although $H_c$ can be reduced to be a few $A \cdot m^{-1}$, their $B_s$ value cannot be increased above 1.95 T because of the microstructural limitations. Here, the $Fe_{68.09}Co_{17.02}B_{10.9}Si_4$ magnetic film has the combination of a silicon-steel like high $B_s$ up to 2.02 T and low $H_c$ of 33 $A \cdot m^{-1}$, challenging the $B_s$-$H_c$ trade-off relationship. As a result, the ultra-high $B_s$ of the $Fe_{68.09}Co_{17.02}B_{10.9}Si_4$ magnetic film is conducive to its application in high-power and size-limited scenarios. Its high resistivity and low $H_c$ are also conducive to the reduction of eddy current loss and hysteresis loss, making it a reliable candidate for high-frequency electronic devices such as thin film inductors.

**Structural origin of unprecedented magnetic properties**

In order to understand the origin of ultrahigh $B_s$, we observed the microstructure of $Fe_{68.09}Co_{17.02}B_{10.9}Si_4$ film using transmission electron microscopy (TEM). The contrast of light-color, dark-color and black-color can be distinguished from bright-field TEM image (Figure 5a), representing different precipitated phases. Through the analysis of its selected area electron diffraction pattern (inserted in Figure 5a), it can be seen that the precipitated nanocrystals are *bcc*-FeCo phase. In order to accurately identify the structures represented by each contrast, Figure 5b shows the high-resolution TEM images near dark areas with increasing magnification. The light-colored areas are typical of the maze-like pattern indicating an amorphous structure. The diffraction halo shown by the fast Fourier transform (FFT) pattern of region (1) also proves the disordered structure (Figure 5e). For the adjacent dark-colored regions, we observe that the pattern of region (2) is still maze-like, but there is a clear interface with the light-colored regions. A diffraction halo was also observed after FFT, but faint diffraction specks were observed nearby (Figure 5f). It indicates that there is a small number of crystal-like regions (CLRs) in the dark region except for the second amorphous structure. These regions seem to be intermediate-range ordered structures of *bcc*-FeCo [84, 85], which is positive to improve the $B_s$ of $Fe_{68.09}Co_{17.02}B_{10.9}Si_4$ film. Figures 5c and 5d show the high-resolution TEM images of black-colored regions with gradual magnification. This "starfish" -like shape is typical of a regularly arranged crystalline structure. The spacing between adjacent crystal planes is 2.008 Å, representing the (110) planes of *bcc*-FeCo. The FFT image (Figure 5g) also proves the existence of crystal structure of the region (3) in Figure 5d.

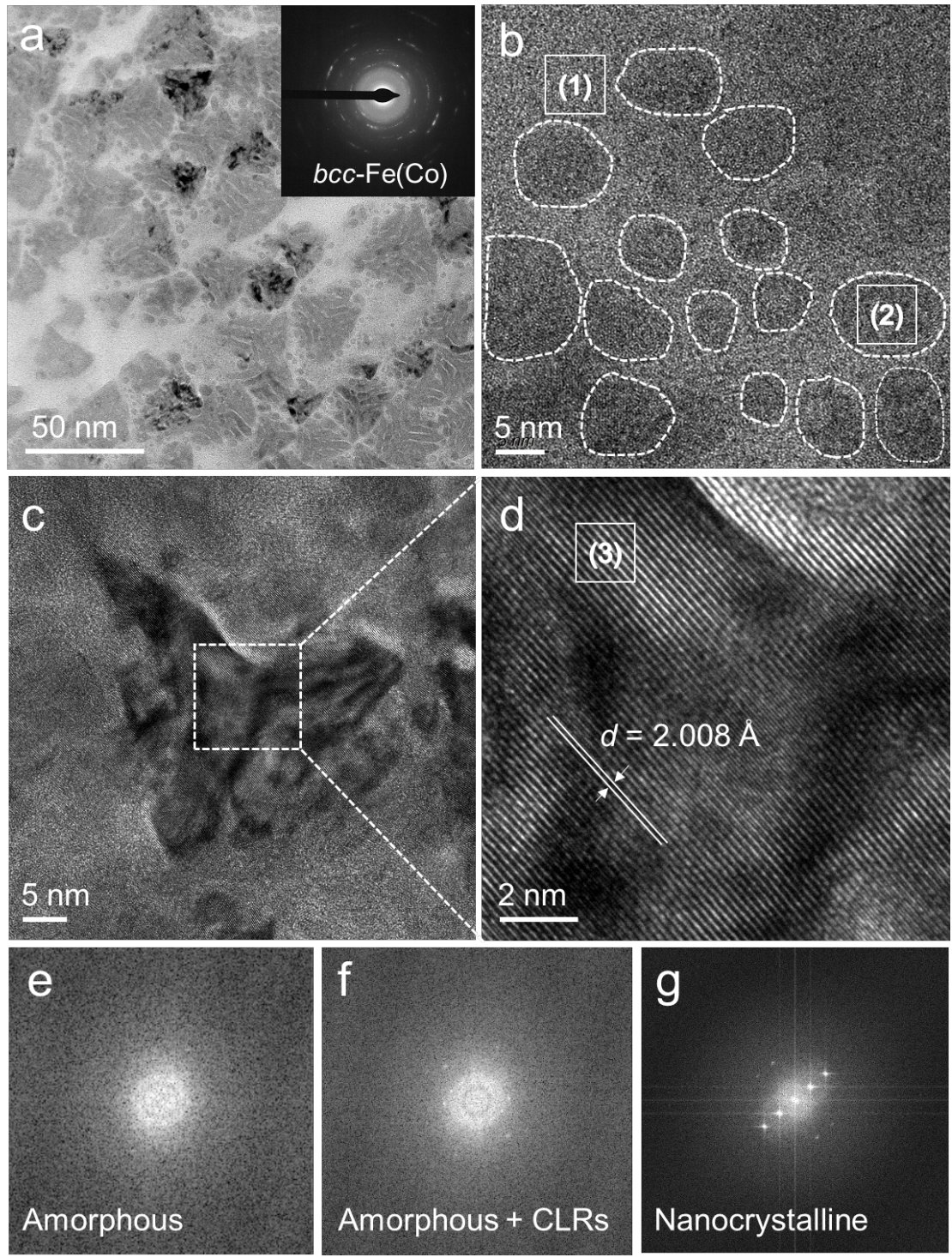

**Figure 5. Microstructure of Fe$_{68.08}$Co$_{17.02}$B$_{10.9}$Si$_4$ soft magnetic film.** (a) The bright-field TEM image of Fe$_{68.08}$Co$_{17.02}$B$_{10.9}$Si$_4$ soft magnetic film, which is distinguished by the contrast of light-color, dark-color and black-color regions. The inset shows a selected area electron diffraction pattern, indicating the precipitation of *bcc*-Fe (Co). (b) The high-resolution TEM images near dark areas. (c, d) The high-resolution TEM images of black-colored regions with gradual magnification. (e-g) The FFT patterns of region (1), region (2) and region (3), respectively.

In order to explain the formation of this biphase amorphous and nanocrystalline composite structure, the distribution of each element was analyzed using energy-dispersive X-ray (EDX) surface scanning technique (Figure S4). The Fe element is enriched in dark-colored and black-colored regions, and little in light-colored amorphous regions. However, Co is only enriched in the black-colored crystal region and low in the dark-colored and light-colored amorphous region. It shows that the clusters sputtered from the same FeCo target still led to the recombination of element distribution and structure in the accumulation. At the location of Fe and Co segregation, magnetic inhomogeneity due to differences in atomic magnetic moments between magnetic elements is more likely to induce crystal formation, corresponding to the black-colored region in Figure 4a. At the dark-colored region without segregation, the amorphous structure is formed due to the enhancement of metalloid elements B and Si on the glass formation ability. The interface between the dark-colored and light-colored regions is caused by the difference in the proportion of magnetic elements in the two phases. Therefore, the precipitation of the second amorphous and nanocrystalline phase with a high proportion of magnetic elements on the amorphous matrix increases the average atomic magnetic moment, greatly increasing the $B_s$ of the film. The biphase amorphous-nanocrystalline structure not only shows excellent comprehensive soft magnetic properties, but also improves the resistivity because of the existence of the amorphous phase, which is beneficial to reduce the energy loss in practical application.

## Conclusion

In summary, we develop a method for the high-throughput screening of soft magnetic properties based on the MOKE technology. The component library was first fabricated by three-target magnetron co-sputtering. Subsequently, the relationship between microstructure and soft magnetic properties was established for the component library through the high-throughput MOKE and XRD. Finally, the $Fe_{68.09}Co_{17.02}B_{10.9}Si_4$ soft magnetic film was successfully developed using this paradigm. The resultant soft magnetic alloy exhibits a combination of ultrahigh $B_s$ of 2.02 T and high resistivity of 882 μΩ·cm. The excellent properties of $Fe_{68.09}Co_{17.02}B_{10.9}Si_4$ soft magnetic film is closely related to the amorphous-nanocrystalline biphase structure, which can maximize the average atomic magnetic moment and complicates the electronic structure. The alloy film developed here is promising for the film inductors of high powder density and high efficiency. The high-throughput technology for screening magnetic properties also provides a paradigm for the development of next-generation soft magnetic materials and understanding the relationship between microstructure and magnetic properties.

## Methods

**Materials fabrication.** The combinatorial library was fabricated by three-target magnetron co-sputtering deposition from elemental sputtering targets, and 100-mm-diameter Si wafers were used as substrates. The ratio of each element in the component library is realized by adjusting the sputtering power of each element target. The code-named I to IV thin films were prepared by single target magnetron sputtering of alloy target with corresponding chemical composition. In order to facilitate the measurement of soft magnetic properties, its substrate was selected to be 1 mm × 1 mm square and 1.8 mm × 60 mm rectangular Si wafers. The thickness of all film samples was ensured to be around 1 um by adjusting the sputtering power and time.

**Structural and compositional characterizations.** X-ray diffractions was used to characterize the structure of the combinatorial libraries and the code-named I to IV thin films by using Bruker D8 X-ray diffractometer with a Cu-$K_\alpha$ radiation source. The measurements on the structure of the combinatorial libraries were performed in an array of rectangular measurement areas, with a distance of 10 mm. The chemical compositions of the code-named I to IV thin films were identified using an inductive coupled plasma emission spectrometer. Transmission electron microscopy observations were performed in a FEI Tecnai G2 F30 S-Twin TEM operated at 300 kV, with the element distribution of the $Fe_{68.09}Co_{17.02}B_{10.9}Si_4$ soft magnetic film was conducted by energy dispersive X-Ray spectroscopy.

**Magneto-optical Kerr effect measurements.** Magnetic domain and magnetic domain motion with the external magnetic field was observed by an em-Kerr-highres magneto-optical Kerr microscope. LED lattice light source is used, and the incidence plane of light source is perpendicular to the direction of magnetization (pure transverse Kerr effect). The amplitude of applied magnetic field parallel to the combinatorial library for measuring all Kerr hysteresis loops was 20 mT. The measurements on the Kerr hysteresis loop of the combinatorial libraries were performed in an array of rectangular measurement areas, with a distance of 5 mm.

The drift and farad corrections of *Kerrlab* software were used to transform the loops uniformly. Some thermal drift of the lamp might lead to a shifted hysteresis loop, notably seen when the starting and end points at equal field value differ in their intensities. With the drift corrections the loop can be closed. If the intensity of the loop increases with rising field beyond nominal saturation of the sample, some Faraday rotation of light in the glass components of the optical system is most likely superimposed to the Kerr signal from the sample. With the farad corrections the loop can be adjusted to display field-independent saturation.

**Properties measurements.** The $B_s$ of the code-named I to IV thin films were measured with a Quantum Design MPMS3 vibrating sample magnetometer under the maximum applied field of 1500 kA/m. Scanning electron microscopy was used to accurately measure the sectional thickness of each 1 mm × 1 mm film with an accuracy of 0.001 um. And $B_s$ (T) is calculated by the following formula:

$$B_s = \mu_0(H + M_s) = \frac{4\pi \times m}{10^5 \times V} + 4\pi \times 10^{-7} \times H$$

where $m$ is the magnetic moment measured by VSM, $V$ is the volume of the film, and $H$ is the external magnetic field applied when magnetization reaches saturation (unit is Oe). This measurement method is accurate because it avoids measuring material density which is prone to error.

Soft magnetic properties including $H_c$, effective permeability and quality factor was measured with a EXPH 100 B-H loop tracer at room temperature and an Agilent 4294A impedance analyzer under the applied field of 1 A/m. The DC resistance of the code-named I to IV thin films were measured by automatic four-point probe method.

# Combinatorial Development of Amorphous/nanocrystalline Biphase Soft Magnetic Alloys with Silicon-steel like Saturated Magnetic Induction

Xuesong Li[1,2 #], Jing Zhou[1,2 #], Xiao Liu[1], Xibei Hou[1], Tengyu Guo[1], Bo Wu[3], Baoan Sun[1,2 *], Weihua Wang[1,2,4], Haiyang Bai[1,2,4 *]

[1]*Songshan Lake Materials Laboratory, Dongguan 523808, China*
[2]*Institute of Physics, Chinese Academy of Sciences, Beijing 100190, China*
[3]*Bay Area Center of Electron Microscopy, Dongguan 523808, China*
[4]*Center of Materials Science and Optoelectronics Engineering, University of Chinese Academy of Sciences, Beijing 100049, China*

**The supporting Information includes:**

**Figures S1-S4.**

[#]These authors contributed equally to this work.

[*]Corresponding authors: sunba@iphy.ac.cn (B. A. Sun); hybai@iphy.ac.cn (H.Y. Bai)

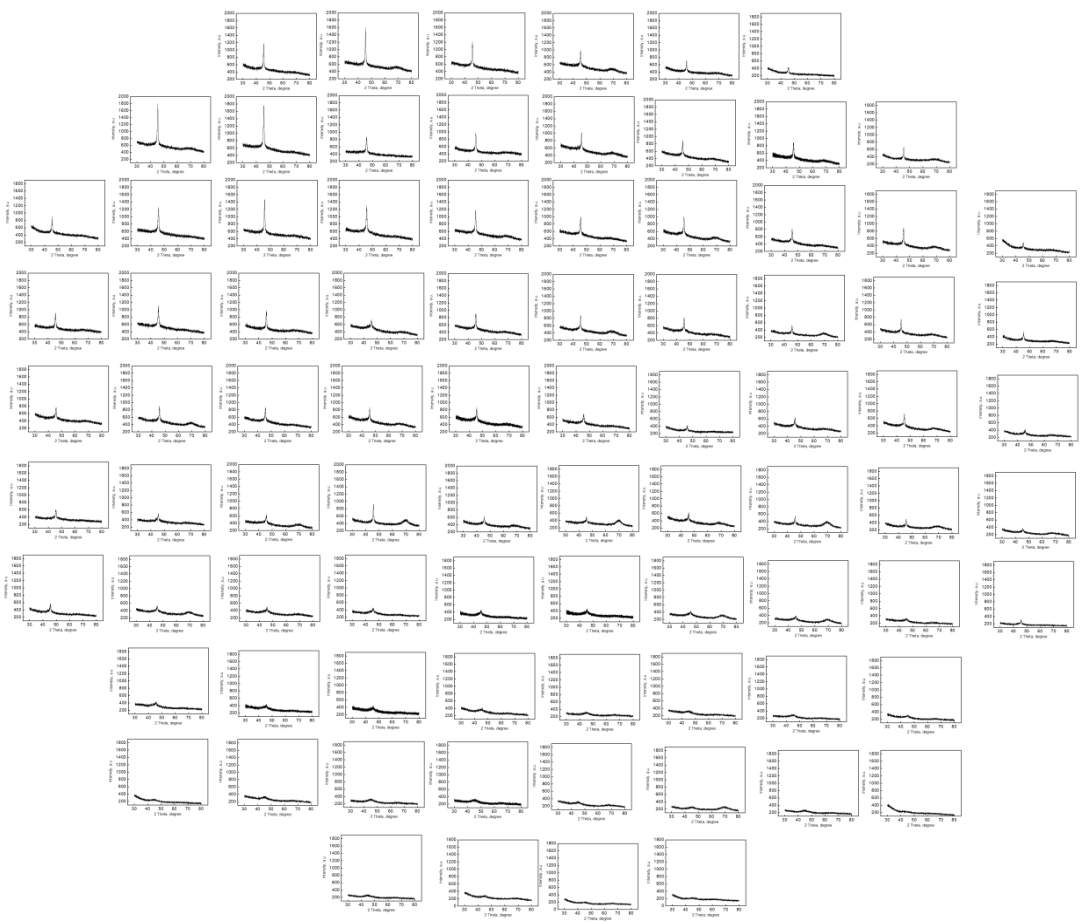

Figure S1. The structure map of the component library, where the spacing between X-ray spots in the horizontal and vertical directions was 10 mm.

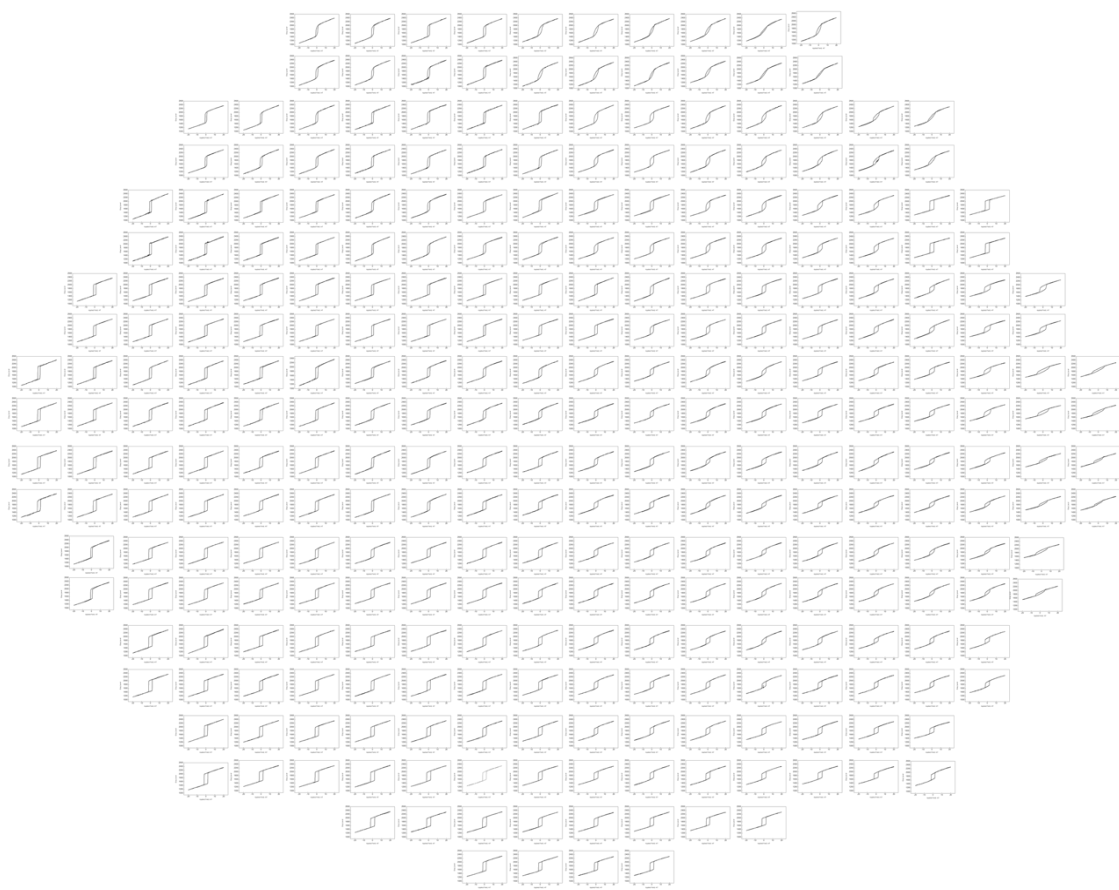

Figure S2. The Kerr hysteresis loops in combinatorial library, where the spacing between test points in the horizontal and vertical directions was 5 mm to ensure efficiency and accuracy.

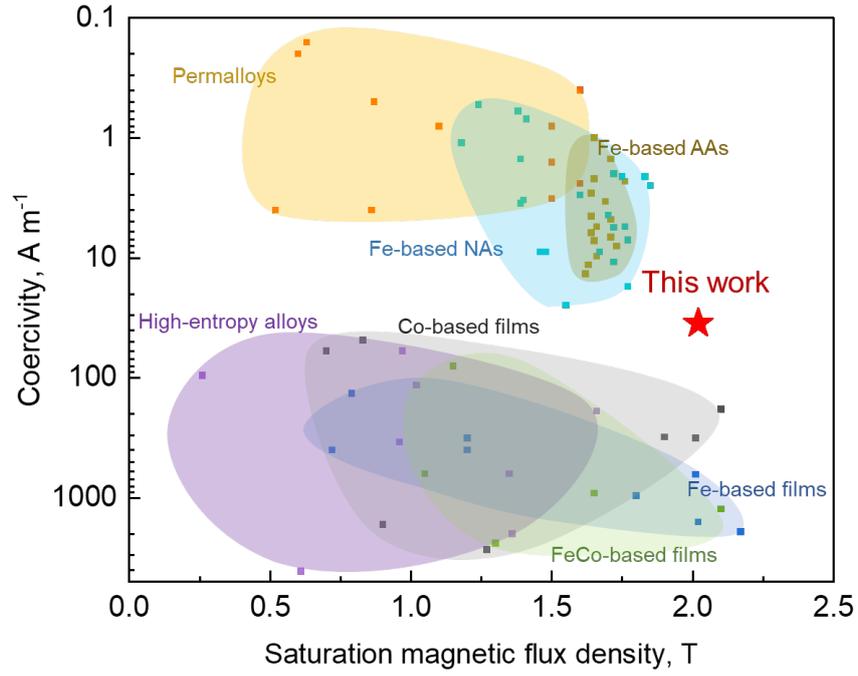

Figure S3. Relationship between the $B_s$ and $H_c$ of the typical soft magnetic alloys and magnetic alloy films.

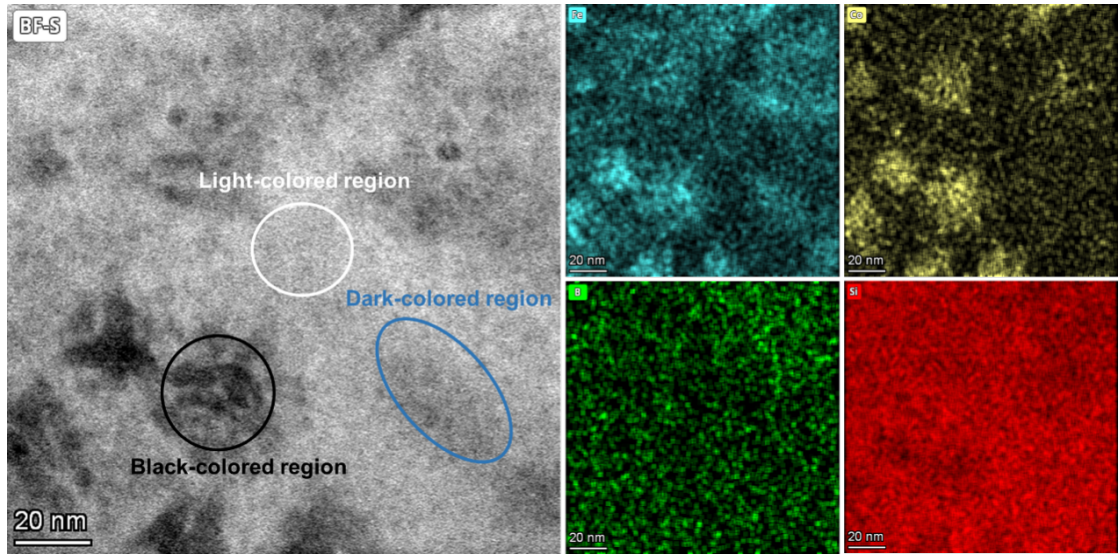

Figure S4. The element distribution of Fe, Co, Si and B using EDS surface scanning technique.